\documentstyle[12pt]{article}
\begin{document}
\begin{center}
{\bf \large
Fluctuation-Stimulated Variable-Range Hopping}

V.I. Kozub

{ \it A.F. Ioffe Physico-Technical Institute, Russian Academy of
Sciences
194021 St. Petersburg, Russia}

S.D.Baranovskii

{\it Institut fur Physikalische Chemie und Zentrum fur
Materialwissenschaften der Philipps-Universitat Marburg, D-35032
Marburg, Germany}

I.Shlimak

{\it  Jack and Pearl Resnick Institute of Advanced technology,
Department of Physics, Bar-Ilan University, Ramat-Gan 52900,
Israel }
\vskip1cm
{\bf Abstract}
\end{center}
Qualitatively new transport mechanism is suggested for hopping
of carriers according to which the variable-range hopping (VRH)
arises from the resonant tunneling between transport states
brought into resonance by Coulomb potentials produced by
surrounding sites with fluctuating occupations. A
semiquantitative description of the hopping transport is given
based on the assumption that fluctuations of energies of hopping
sites have spectral density $1/f$.

\vskip1cm

Recently, it was suggested that fluctuations of site occupations
in thermal equilibrium can cause essential energy fluctuations
for sites participating in the hopping transport [1,2]. This
idea was used to explain the low-frequency noise with spectral
density $1/f$ reported for hopping conductivity in p-Ge(Ga) and
n-Ge(As) [1]. Due to the long-range character of the Coulomb
potential, fluctuations in occupation numbers of some sites lead
to essential fluctuations of energies of the surrounding sites.
In Refs.[1,2], an attempt has been done to distinguish between the
sites participating in the hopping transport (transport sites)
and fluctuators, the latter being the sites which produce energy
fluctuations on the transport sites, but do not directly
participate in hopping processes. Comparison of experimental
data with simple theoretical estimates [2] shows that the noise
level is so large that each effective resistor on the transport
cluster has relative fluctuations of the resistivity of the
order unity. In the present report we study the influence of
such fluctuations on the variable-range hopping.

According to
the theoretical picture suggested in Refs.[1,2], the current noise
is due to fluctuations of site energies affecting the hopping
probabilities. The fluctuations of site energies are caused by
electron transitions between sites which do not belong to the
percolation cluster, but can influence the sites on the cluster
by their Coulomb potentials. Thus this noise is a direct
manifestation of the sequential Coulomb correlations in a
hopping system suggested first by Pollak and Knotek [3,4].  The
effect of these correlations on hopping transport has been
discussed in numerous papers, although most studies in this
field were based so far on computer simulations.  Among very
interesting results one can mention those of the recent study of
Perez-Garrido et al. [5], who showed that the sequential
correlations along with electron polarons have significant
influence on the hopping conductivity in the Coulomb gap regime.
Much less number of attempts have been made yet to develop
analytical approaches to study the influence of sequential
correlations on hopping conductivity.  Ortuno and Pollak [6]
were the first who considered analytically the influence of
these correlations on the activation energy of hopping
conductivity under the assumption of a sharp Coulomb gap. Using
the mean-field theory they have shown that sequential
correlations can reduce the width of the Coulomb gap $\Delta_C$ and
hence diminish the activation energy of hopping conductivity.

In the present report, we study analytically the effect of
sequential correlations within a somewhat different approach.
Following the ideas of Refs.[1,2], we presume, that sequential
correlations cause fluctuations of site energies of the
effective sites on the percolation cluster with spectral density
$1/f$. In general case it is difficult, if possible at all, to
distinguish between transport sites and fluctuators [7] and we
are not able to perform this distinguishing in the whole
frequency range. However, in the limit of the high-frequency
noise it is reasonable to assume [1,2] that fluctuators are
represented by compact pairs of sites each pair having one
electron with correspondingly high transition rate between the
sites of the pair.
Such pairs play the role of electronic two-level systems
[2] and, as the well-known two level systems in structural
glasses, lead to $1/f$ spectrum of fluctuations. Thus in this
frequency region the fluctuations of energies of hopping sites
are caused by hopping ransitions of electrons within
separated low-energy
pairs surrounding the hopping sites similar to those considered
by Efros et al. [8]. Below we will assume that this $1/f$
fluctuations spectrum
starting from the cut-off frequencies within the high-frequency
region mentioned above is extended (with the same parameters)
to much lower frequencies
where the fluctuations are dominated by the aggregates of
defects more complex than the isolated pairs mentioned above.
Having information neither about exact picture of these aggregates
nor about the character of the corresponding relaxational modes,
we still believe that they support the same $1/f$ fluctuation
spectrum started at high frequencies where it is supported by
isolated pairs.
We would like to emphasise that this assumption is decisive
for the whole consideration suggested below. We cannot justify
this assumption for the broad frequency range and we take it
just as an ansatz.
We also assume
the fluctuations of site energies in the vicinity of the Fermi
level to be so large that the energy positions of electrons on
hopping sites within some energy strap of width $\Delta <
\Delta_C$ can be
arbitrary with respect to the Fermi level. In such a regime, the
role of quantum mechanical resonant tunneling becomes important
leading to the universal temperature-independent preexponential
factor in the expression for the hopping conductivity.

An
attempt to draw a picture of the phononless variable-range
hopping based on the resonant tunneling has been recently made
by Baranovskii and Shlimak [9]. They calculated resonant
transition probabilities in the regime of strong fluctuations of
site energies. They also assumed that sites with energies in the
vicinity of the Fermi level have equilibrium occupations.
However, the latter assumption seems not consistent with the
picture of strong energy fluctuations. In the present report we
try to derive the variable-range hopping transport mechanism
which is not based on the equilibrium occupations.

Our main
idea is the following. Due to electron transitions in the
"fluctuators", the energies of sites in the vicinity of the
Fermi level, i.e., of sites responsible for hopping transport at
low temperatures, perform strong energy fluctuations within some
effective range $\Delta$. The energy range $\Delta$
increases with increasing
fluctuation amplitudes. Sites outside this energy range posses
equilibrium occupations.  Transport is due to resonant
transitions which occur when the fluctuating energy of a filled
site within the range $\Delta$ coincides with the fluctuating energy of
an empty site in this range. The critical hopping length $r_c$ is
determined by the solution of the geometrical percolation
problem on random sites with concentration $N_{\Delta}$,
the latter being
the concentration of sites in the energy strap $\Delta$ in the vicinity
of the Fermi level. The amplitude of the site energy
fluctuations determining $\Delta$ depends on temperature $T$
and it is
via $\Delta$ how the temperature influences the conductivity. With
rising $T$ the width $\Delta$ increases leading to the increase of the
concentration $N_{\Delta}$ of effective sites which participate in the
resonant-tunneling processes. Increasing $N_{\Delta}$ leads to decreasing
$r_c$. This can be well called "the variable-range resonant
tunneling" [9], because the higher is $T$, the shorter in space
are effective hops. Below we give arguments which lead us to
such picture of hopping transport.

Let us consider a pair of
sites with site 1 initially occupied and site 2 initially empty.
Let the activation energy for electron transition between the
sites be E and the distance between the sites be $R$. Transition
rate of an electron from site 1 to site 2 has the form
\begin{equation}
\nu_{1,2} = \nu_0 \exp (-2R/a - E/T)
\end{equation}
where $\nu_0$ is the attempt to escape frequency, $a$ is the localisation
length and temperature is measured in the units of energy. In the
absence of fluctuations of site energies the typical transition time
$t_{1,2}$ is determined by the condition $t_{1,2} \simeq
\nu_{1,2}^{-1} \equiv t_c$. The
fluctuations of the site energies can diminish the activation energy
of such a pair by some amount $\delta E$. Given some spectral density of
fluctuations, the quantity $\delta E$ is time dependent.
We assume here the
spectral density of fluctuations in the form $<(\delta E)^2> =
\alpha^2(f_0/f)$; $f < f_0$
where $f$ is the frequency and $\alpha$, $f_0$
are some parameters.

We assume that
fluctuations originate mainly due to pair excitations with
nearly constant density of states [8]. The strongly interacting
pairs form a sort of "dipolar Coulomb glass". Thus one expects
that the $1/f$ spectrum extends up to very low frequencies,
particularly due to the nearly-degenerate character of the
ground state of the Coulomb glass [10]. The high-frequency
cut-off $f_0$ should be related to the fastest possible modes of
the dipolar glass related to hops within single pairs
independent from the rest of the glass. Correspondingly, the
value of $\alpha$ can be related to the magnitude of fluctuations
originated due to these single-pair hops. We start from a
phenomenological picture not specifying the values of $\alpha$
and $f_0$
and having in mind to return to this problem later.

With an account of such fluctuations the rate equation for electron
transition within the chosen pair has the form
\begin{equation}
{\rm d}n_2/{\rm d}t = \nu_0 exp \left(-2R/a - (E + \delta E(t))/ T
\right)n_1 ,
\end{equation}
where occupation numbers $n_2$ and $n_1$  have
initial values
$n_1(0) = 1$;
$n_2(0) = 0$. This equation has solution
\begin{equation}
 n_2(t) = \int_0^t {\rm d}t' \nu_0 \exp \left(-2R/a -
(E+ \delta E(t'))/T \right)n_1(t') .
\end{equation}
Transition time $t$ is approximately determined by the
condition
\begin{equation}
\int_0^t {\rm d}t' \nu_0 \exp \left(-2R/a - (E+\delta E(t'))/T
\right) \sim 1 .
\end{equation}

The integral is obviously dominated by the smallest possible value of
$\delta E(t)$. We replace the term
$\delta E(t)$ in the
integrand by the
mean quadratic fluctuation of the activation energy
$-|<\delta E(t)>|$.
Using Eqs.(2), (5), one obtains for the typical transition
time $t$
\begin{equation}
\alpha \left(\ln (tf_0)\right)^{1/2} \sim T \ln (t_c/t) .
\end{equation}
Of course, the solution of this equation is related to the typical
transition time only in the case that the latter is much longer than
the time of the pure tunnelling without activation $t_h = \nu_0^{-1}
\exp
(2R/a)$. Combining this restriction with the formal solution of
Eq.(6) one obtains
\begin{equation}
\ln (tf_0) = \max \left( \ln(t_hf_0);
\ln (t_cf_0)- \lbrack (\alpha^2/2T^2)^2 + (\alpha/T)^2\ln(t_cf_0
\rbrack^{1/2} +
(\alpha^2/2T^2)^2\right).
\end{equation}

In the case $(\alpha^2/2T^2)^2 <
(\alpha/T)^2(\ln (t_cf_0))^{1/2}$ this result reflects a
decrease of the effective transition time $t$ with respect to its value
$t_c$ in the absence of fluctuations caused by the reduction
of the activation energy. In that sense our result is similar to that
obtained by Ortuno and Pollak [6]. If the parameters of a system
under study correspond to the opposite case
$\alpha^2/4T^2 >
\ln (t_cf_0) $
one can use a series expansion in Eq.(6) which
leads to the expression
\begin{equation}
\ln (tf_0) = \max \left( \ln (t_hf_0); (\ln(t_cf_0))^2T^2/\alpha^2 \right).
\end{equation}
Eq.(7) shows that the transition time in the considered pair
corresponds to the maximal of the two times: the time of pure
tunnelling and the time necessary to nullify the activation energy
for the transition.
This scenario is different from the standard VRH where
the logarithms of the tunneling term and the activation terms
should be compared rather than the corresponding times.
We assume below that it is condition (7) that governs transition
times in the pairs essential for hopping transport in the
fluctuation regime.

In order to
evaluate the hopping conductivity using given transition rates,
one has to formulate the binding criterion for construction of
the percolation cluster. Let us define the binding criterion by
the condition that the time of the pure tunneling $t_h$ in a pair
is of the order of the time necessary to eliminate the
activation energy $E$ of the pair. Herewith we refuse to estimate
numerical constants in the exponential terms for the hopping
conductivity, though we believe that our criterion gives a
correct set of parameters in the exponents in analogy to the
standard variable-range hopping approach. Our criterion
corresponds to the optimization procedure that leads to
approximate equality of two terms in the right-hand side of
Eq.(7).

In order to determine the width $\Delta$ of the energy strip
in which electron transitions are responsible for transport, let
us consider the density of states (DOS) function. We assume that
DOS at $T=0$ is determined by the soft Coulomb gap with quadratic
energy dependence [11]:
\begin{equation}
g(\varepsilon) = g_0 \varepsilon^2 ,
\end{equation}
where $\varepsilon$ is the site energy measured
with respect to the Fermi level. Its integral up to some energy
$\varepsilon$ determines the concentration of cites
$N_{\varepsilon} = (2/3) g_0 \varepsilon^3 $ with
energies less than $\varepsilon$. Thus one estimates as $R_{\varepsilon}
= c N_{\varepsilon}^{-1/3}$ the critical distance between the sites
within the percolation cluster involving the sites from this region
of energies ($c \simeq 0.87  $, [11]).

 Using this estimate
for the first term in the right-hand side of Eq.(7) and
neglecting logarithmic terms, one obtains that energy ( that
provides equality between two term in the r.h.s. of Eq.(7) is
determined by the relation
$\ln (t_hf_0) \sim  (4/3)c g_0^{-1/3}\Delta^{-1}/a$.
In
the same way one estimates for the second term in the right-hand
side of Eq.(7) $\ln(t_cf_0) \sim \Delta/T$.
Thus from the binding criterion
one has
\begin{equation}
\Delta \simeq  (4c g_0^{-1/3}\alpha^2/3a)^{1/3} .
\end{equation}
It is worth noting that except of a numerical factor the same estimate
is valid for 2D case.

According to the transport
picture formulated above, the hopping conductivity is determined
by the solution of the geometrical percolation problem on
randomly distributed sites with concentration $N_{\Delta}$. The width of
the energy strip $\Delta$ depends on temperature via the temperature
dependence of parameter $\alpha$. We will discuss this dependence
later. Before doing so we would like to justify our transport
picture by showing that in contrary to standard VRH approaches,
in the suggested transport mechanism the occupation numbers of
sites which form the transport path are independent on
temperature and hence these occupation numbers are not essential
for the calculation of the temperature dependence of the hopping
conductivity.

Taking into account the explicit time dependence
of the site energies one can write the rate equation for the
occupation number of a site $i$ in the form
\begin{equation}
\frac{{\rm d} n_i}{{\rm d} t} = \sum_{j} \left( W_{j,i}(\varepsilon
_i(t), \varepsilon_j(t))n_j(1-n_i) -
W_{i,j}(\varepsilon
_i(t), \varepsilon_j(t))  n_i(1-n_j)\right)
\end{equation}
where for  hopping probabilities one has $\max W_{i,j} =
\max W_{j,i} = \nu_0 \exp (-2 R_{i,j}/a)$.
The time dependence of the probabilities does not allow to
apply a simple detailed balance considerations. Moreover,
the 1/f spectrum of the fluctuations leads to the dependence
$|\delta \varepsilon| \propto \ln^{1/2}(tf_0)$ which does not allow
any time averaging between the hopping events. This is a clear
manifestation of non-ergodic behavior.
The electron hop is prepared by the surrounding which implies
memory effects instead of ergodicity.

Energy fluctuations during the typical hopping time do not allow
to specify any ordering of states with respect to their energies
within the energy strip $\Delta$ in the vicinity of the Fermi level. As
a result, occupation numbers of those states do not correspond
to the Fermi distribution being rather arbitrary. For energies
outside the range $\Delta$ , fluctuations cannot suppress the energy
ordering of the sites. According to the very idea of the above
estimation of the hopping time $t_{\Delta}$,
the latter corresponds to the
relaxation time of a charge in the energy strip of the width $\Delta$
(determined by Eq.(10)) and thus it characterizes the largest
possible time for the charge transfer within the system. Having
this in mind, one expects that the ergodicity and detailed
balance considerations are restored at time scales larger than
$t_{\Delta}$ and correspondingly for energy scales larger than
$\Delta$.  Thus
one obtains for the time-averaged occupation number of a state $i$

\begin{eqnarray}
\sum_j ( <W(\varepsilon_j, \varepsilon_i><n(\varepsilon_j)>
(1 - <n(\varepsilon_i)>) - <W(\varepsilon_i,
\varepsilon_j)>\cdot \nonumber\\
<n(\varepsilon_i)> (1 - <n (\varepsilon_j)>) ) = 0
\end{eqnarray}
where $<>$ denotes the time average. For the
probabilities $<W(\varepsilon_i, \varepsilon_j)>$
one obviously has ${\rm d} <W>/{\rm d} <\varepsilon_{i,j}>|_{
\varepsilon_{i,j} \rightarrow 0} \leq
<W>/\Delta$ due to averaging over instantaneous values of the
energies within the relevant interval $\sim \Delta$.
Thus the
the averaged occupation numbers $<n_i>$ near the Fermi level
correspond to a smeared Fermi distribution with effective
temperature of the order of $\Delta$.
For energies $|\varepsilon| >> \Delta$ the
density of states is larger, the tunneling time is smaller and
the distribution tends to the thermal one.

The fact that site
occupation numbers near the Fermi level are arbitrary justifies
the procedure applied above where we neglected the occupation
number factors restricting ourselves to the diffusion
consideration. It is also worth noting that according to the
above picture, the arguments leading to the Coulomb gap [8] do
not hold anymore implying that the Coulomb gap is suppressed in
the energy region of the width $\Delta$ in the vicinity of the Fermi
level.

In our model, hopping transitions correspond to
processes in which activation energy is eliminated by
fluctuations of the site energies due to correlated electron
transitions in the surrounding pairs. In such a case, special
attention should be devoted to resonant phononless electron
transitions. Until now, we mainly discussed a preparation of the
system of two sites to the electron hop rather than the hop
itself. One should realize that fluctuations with spectral
density $1/f$ contain contributions of all possible modes with
frequencies in the range between $f_0$ and $1/t$, where $t$ is the
hopping time. Function $\ln^{1/2}(tf_0)$ represents a sort of "envelope
function" for the fluctuation amplitudes and activation energy
for a hop can vanish many times before the hop really occurs.
 For a pair of hopping sites 1 and 2 with energies
$\varepsilon_1$
and $\varepsilon_2$, the routine perturbation theory
[13] under adiabatic
conditions gives the probability $p$ of the resonant transition during
a "fluctuation period" $p \sim I^2/\hbar |(dE/dt)|^{-1}$, where
$I$ is the overlap
integral $ I = I_0\exp(-2R/a) $ and  $E = \varepsilon_1 -
\varepsilon_2$.
Making use of the random
character of fluctuations and hence summing the probabilities for
different "periods", one obtains for transition rate between sites 1
and 2
\begin{equation}
W \sim  I^2/Eh .
\end{equation}

Assume that our sites 1 and 2 form the effective resistor on the
percolation cluster. Under applied external voltage V, dc
current through the chosen resistor is equal to
\begin{equation}
j = eW<n_1(\varepsilon_1(t)+eV,t) - n_2(
\varepsilon_2(t),t) > \sim  e^2WV {\rm d}<n(\varepsilon)>/{\rm
d}\varepsilon,
\end{equation}
where $W$ is determined by Eq.(13).
Taking into account the estimate ${\rm d}<n>
/{\rm d}\varepsilon \sim 1/\Delta$ we come to the
following expression for the hopping current
\begin{equation}
j \simeq (e^2/h) V I^2/\Delta^2 .
\end{equation}
The distance $R$ in the exponent of the overlap
integral $I$ is controlled by the geometrical binding criterion
discussed above.

Preexponential term $I_0$ of the overlap
integral depends on the character of the hopping sites.
For sites with the hydrogen-like Coulomb potentials
$I_0 = (2e^2/3 \kappa a)(r/a)$, $r$ being characteristic hopping
length, $\kappa$ being the dielectric constant. In this case
$I_0 /\Delta \sim (r/a)^2 >> 1$.
Different situation can appear for sites with screened
Coulomb potential. In the case of impurity
screening  $I_0 \sim e^2/\kappa r \sim \Delta$.
Using this estimate for $I_0$ and the estimate for the
exponential term in the overlap integral from Eqs.(7),(9)
one obtains for the hopping conductance
\begin{equation}
G \simeq (e^2/h) \exp(- (T'/T)^{\beta};\hskip0.5cm
(T'/T)^{\beta} = (4c/3g_0^{1/3}a\alpha)^{2/3}
\end{equation}
It can look really curious that we pretend to determine
preexponential factor in the expression for hoping conductivity,
while the very derivation of the exponent based on the
order-of-magnitude binding criterion cannot provide us with a
correct numerical factor in the exponent. However, our
derivation of the preexponential factor in Eq.(15) seems to
represent just a revision of the relation between diffusivity
and mobility (Einstein relation) in the non-ergodic situation
caused by correlated hops with $1/f$ spectrum. The crucial
estimate for the above derivation is the approximate equality
between $I_0$ and $\Delta$, both being controlled by the Coulomb energy.
The situation is to some extent similar to the standard band
conduction via extended states in a Fermi system with a single
characteristic energy equal to the Fermi energy.

In order to
derive the values of $T'$ and $\beta$ it is necessary to specify the
microscopic picture of the fluctuations which lead to the
resonant transitions. Following ideas of Ref.2, we assume that
the crucial parameter of the fluctuation spectrum $\alpha$ can be
related to the fluctuation potential caused on a hopping site by
the nearest soft pair with the activation energy $\sim T$. This gives
the estimate $\alpha  = A(\Delta_CT^2)1/3$, where $A$
is a numerical coefficient
[2, 12]. Substituting this expression for $\alpha$ into Eq.(15), one
obtains
\begin{equation}
\beta = 4/9, \hskip1cm T' = (4 c/3Aag_0^{1/3}\Delta_C^{1/3})^{3/2}.
\end{equation}
Note that the value $\beta = 4/9$ can hardly be distingwished
experimentally from the "standard" value $\beta = 1/2$ for
Efros-Shklovskii law. It can be also mentioned that for $A> 1$
the temperature $T'$ appears to be smaller than the characteristic
temperature $T_1$ for Efros-Shklovskii law; this fact corresponds
to stimulation of the hopping by fluctuations (additional effect
is related to an increase of the pre-exponential mentioned above).

The above estimate of $\alpha$ was based on the parabolic Coulomb gap
(Eq.(8)) which is believed to exist in 3D systems. For 2D
systems, pair excitations are known to be less important [11].
Nevertheless, we believe that our results can be also applied to
many situations where transport takes place in a 2D layer
surrounded by a 3D system of impurities. Charge hops within the
3D dopant system can give rise to the effective energy
fluctuations on transport sites leading to the transport
mechanism suggested in this report. Therefore we think that
Eqs.(15),(16) might provide a theoretical explanation for
recent experimental data that show in some cases a universal
prefactor $e^2/h$ in the exponential temperature dependence of the
2D hopping conductivity at low temperatures [12]. More study is
needed, of course, to clarify this question.
\vskip0.5cm

{\large Acknowledgments.}
\vskip0.5cm
Authors are grateful to M. Pollak for useful
discussions. Financial support of the
Deutscheforschungsgemeinschaft via SFB 383 and that of the Fonds
der Chemischen Industrie is gratefully acknowledged. V.I.K.
acknowledges a support by Russian Foundation for Fundamental
Research under the Grant N 97-02-18280a.

\vskip1cm

{\bf \large References}
\vskip0.5cm
1. I. Shlimak, Y. Kraftmakher, R. Ussyskin, and K. Zilberberg,
Solid State Commun. {\bf 93}, 829 (1995).

2. V.I. Kozub, Solid State
Commun. {\bf 97}, 843 (1996).

3. M. Pollak, Disc. Faraday Soc. {\bf 50}, 11
(1970).

4. M.L. Knotek and M. Pollak, J. Non-Cryst. Solids {\bf 8-10},
505 (1972); Phys. Rev. B 9, 644 (1974).

5. A. Perez-Garrido, M.
Ortuno, E. Cuevas, J. Ruiz, and M. Pollak, Phys. Rev. B {\bf 55},
R8630 (1997).

6. M. Ortuno and M. Pollak, J.Phys.C: Solid State
Phys. {\bf 16}, 1459 (1983).

7. M. Pollak, Philos. Mag. B {\bf 8}, 535
(1984); M. Pollak, private communication (1999).

8.  A.L. Efros,
J.Phys.C: Solid State Phys. {\bf 9}, 2021 (1976); S.D. Baranovskii,
A.L. Efros, B.I. Shklovskii, Sov. Phys. JETP {\bf 51}, 199
(1980).

9. S.D. Baranovskii and I. Shlimak, cond-mat./9810363 (1998).

10. Sh.M. Kogan, Bull. Am. Phys. Soc. 42, 777 (1997).

11. B.I.
Shklovskii and A.L. Efros, "Electronic Properties of Doped
Semiconductors" (Springer, Berlin, 1984).

12. N.V. Agrinskaya and
V.I. Kozub, Physica status solidi {\bf 205}, 13 (1997).

13. L.D.
Landau and E.M. Lifshits, Quantum Mechanics, (Moscow, 1974, in
Russian) p. 174.

14. W. Mason, S.V. Kravchenko, G.E. Bowker, and
J.E. Furneaux, Phys. Rev. B {\bf 52}, 7857 (1995); S.I. Khonduker,
I.S. Shlimak, J.T. Nicholls, M. Pepper, and D.A. Ritchie, Phys.
Rev. B {\bf 59}, 4580 (1999).

\end{document}